# Twist Engineering for Reconfigurable Optical and Optoelectronic Devices

Gang Huang[1†], Annan Helian[1†], Mengting Jiang[1*], Chi Wang[2], Yu Xing[1], Mayank Joshi[1,3], Jae Yeong Lee[1], Jiang Wang[4], Syed M Assad[1], Ping Koy Lam[1,5], Qiushi Liu[6*], Lin Wu[2*], Young-Wook Cho[1,5*], Yuan Ma[7*], Xuezhi Ma[1,5*]

1. Quantum Innovation Centre (Q.InC), Agency for Science, Technology and Research (A*STAR), 4 Fusionopolis Way, Kinesis, #05, 138635, Singapore.

2. Science, Mathematics, and Technology (SMT), Singapore University of Technology and Design (SUTD), 8 Somapah Road, 487372, Singapore

3. Australian National University (ANU) Department of Quantum Science and Technology, Australian National University, Canberra, ACT, 2601, Australia.

4. Hangzhou Institute for Advanced Study, University of Chinese Academy of Sciences, Hangzhou, China

5. Centre for Quantum Technologies (CQT), National University of Singapore (NUS), 117543, Singapore.

6. Shanghai Institute of Optics and Fine Mechanics, Chinese Academy of Sciences, Shanghai, 201800, China

7. Department of Mechanical Engineering, The Hong Kong Polytechnic University, Hung Hom, Hong Kong S.A.R., China

Corresponding authors email address:
Mengting Jiang: mengting_jiang@a-star.edu.sg;
Qiushi Liu: liuqiushi@siom.ac.cn;
Lin Wu: lin_wu@sutd.edu.sg;
Young-Wook Cho: cho_youngwook@a-star.edu.sg;
Yuan Ma: y.ma@polyu.edu.hk;
Xuezhi Ma: ma_xuezhi@a-star.edu.sg;

**Abstract**

Reconfigurable optical and optoelectronic devices require compact tuning mechanisms capable of reshaping electronic, excitonic, polaritonic, and photonic responses without rebuilding the underlying nanostructure. Against this backdrop, twist has emerged as a powerful geometric degree of freedom that reconfigures interlayer coupling, momentum matching, symmetry, radiation channels, and chiral response by simply rotating adjacent two-dimensional layers or photonic lattices. In this Review, we survey twist-engineered optical and optoelectronic devices spanning van der Waals materials and photonic platforms. We first review the current landscape of twist-angle metrology, classifying existing characterization approaches into three complementary categories: direct structural imaging, methods based on moiré periodicity and morphological features, and techniques that infer the twist angle from spectroscopic or electronic responses. We then survey the principal technological routes for twist-angle control, including deterministic transfer and growth strategies, atomic force microscopy (AFM)-assisted manipulation, quantum twisting microscopy (QTM), microelectromechanical systems (MEMS), and emerging non-contact approaches, highlighting their respective capabilities, limitations, and prospects for programmable and scalable moiré photonic platforms. Finally, we discuss the future evolution of twist engineering from the fabrication of individual twisted structures toward dynamically reconfigurable, feedback-controlled, and manufacturable photonic systems. We further highlight MEMS-based rotation, piezoelectric actuation, and micro-LiDAR as representative enabling technologies and emerging applications within this broader landscape. We anticipate that twist engineering will evolve from a geometrical design concept into a universal platform for realizing adaptive, multifunctional, and programmable optical and optoelectronic systems.

## 1. Introduction

Reconfigurable devices are becoming a central requirement in nanophotonics[1, 2], optoelectronics[3-6], nonlinear optics[7-13], quantum optics[14-25], integrated sensing and mechanics[26-28]. But many emerging systems require responses that can be tuned after fabrication, including spectral selectivity, polarization, emission direction, nonlinear conversion, carrier transport, and light-matter coupling.

However, achieving efficient, compact, and broadly reconfigurable optical functionality remains a fundamental challenge. Conventional reconfiguration strategies rely on carrier injection, thermal tuning, phase-change materials, liquid crystals, electro-optic modulation, mechanical scanning, or post-fabrication trimming[29-32]. These approaches have enabled important progress, but they often involve trade-offs among optical loss, tuning range, speed, reversibility, footprint, and process compatibility[33]. In parallel, metasurfaces and photonic crystals have made wavefront engineering compact, yet most high-performance implementations remain either static or only weakly tunable[34-37].

Twist offers a distinct route to reconfiguration because it changes the relationship between two layers rather than the composition of either layer. In two-dimensional materials, rotational misalignment creates moiré superlattices that reshape electronic bands, excitons, phonons, polaritons, valley selection rules, and interlayer transport[38-40]. In photonic structures, twisting two lattices reconstructs reciprocal-space coupling, guided resonances, flat bands, BIC-related radiation channels, and far-field chirality[41-43]. Across both families, twist (or twist angle) becomes an active design variable.

This characteristic is especially versatile because twist engineering can, in principle, be static, discretely reset, continuously tuned, or coupled to other degrees of freedom such as interlayer spacing, strain, lateral displacement, and compression[26, 28, 39, 44]. The same concept therefore links deterministic tear-and-stack fabrication, in situ probe manipulation, quantum twisting microscopy, strain programming, MEMS-enabled rotation, piezoelectric actuation, and emerging contactless approaches. The key question is no longer only how to make a twisted structure, but how to control twist with the precision, stability, speed, and scalability required by devices.

Here we review reconfigurable twist-engineered two-dimensional and photonic devices. We first introduce twist as a device-level control parameter, then summarize the metrology and fabrication tools needed to quantify and implement it. We then discuss two device classes: twisted two-dimensional material systems, where twist modifies electronic, excitonic, polaritonic, and optoelectronic responses; and twisted photonic systems, where twist controls resonances, radiation, polarization, chirality, nonlinear emission, wavefronts, sensing, imaging, and compact beam steering. This framing keeps MEMS and micro-LiDAR as important examples while placing them within the broader field of programmable twist-enabled devices.

By organizing the field around device functionality, this Review aims to identify the common design rules, bottlenecks, and integration opportunities that connect twistronics, moiré photonics, adaptive metasurfaces, and chip-scale optical systems.

## 2. Metrology and Characterization of Twist-Engineered Two-Dimensional Materials

For reconfigurable devices, the twist angle is more than a structural descriptor; it is a fundamental design variable that governs the moiré length scale[38], local atomic registry[45], interlayer hybridization[46], and symmetry of the coupled system[26, 39, 45, 47]. Its effects are further coupled to interlayer spacing, strain, lattice relaxation, and environmental boundary conditions, collectively determining the optical and electronic response of the device[44]. From a device perspective, twist angle must therefore be precisely measured, controlled, and maintained, as its value, spatial uniformity, temporal stability, and measurement accuracy directly influence device reproducibility and performance. Existing twist-angle characterization techniques can be broadly classified into three categories: direct structural characterization[48-50], methods based on moiré periodicity and morphological feature[51-54], and methods based on spectroscopic or electronic responses[55-57].

### *2.1 Direct Structural Characterization*

The first class comprises methods based on **direct structural characterization**, whose central principle is to resolve the relative crystallographic orientations of the constituent layers in either real or reciprocal space, thereby extracting the twist angle under minimal model assumptions. Representative examples include scanning transmission electron microscopy (STEM) and its associated diffraction-based techniques[48, 49]. In STEM imaging, particularly under annular dark-field (ADF) or low-angle annular dark-field (LAADF) modes, twisted bilayer or multilayer two-dimensional crystals exhibit moiré superlattice patterns with well-defined periodicity in real space[50]. By performing Fourier transforms of atomically resolved STEM images, reciprocal lattice vectors from individual layers can be simultaneously identified, and their relative rotational offset directly corresponds to the interlayer twist angle[58].

Selected-area electron diffraction (SAED) provides direct access to the reciprocal-space structure of crystalline lattices. In STEM, a highly convergent electron probe with a lateral size on the order of ~1 Å is employed to interrogate the atomic structure of a crystal. As the electron beam traverses a sufficiently thin specimen, periodic scattering from lattice planes gives rise to diffraction spots whose positions are intimately linked to the lattice spacing and crystallographic orientation. In twisted two-dimensional materials, such as twisted bilayer graphene, the diffraction patterns arising from the two constituent monolayers appear as two rotated sets of reciprocal lattice vectors. By analyzing the angular offset between these diffraction spots, the relative crystallographic orientation between the layers, namely the interlayer twist angle, can be directly extracted[57, 58]. These approaches enable direct resolution of atomic lattices or moiré superlattice structures. However, they typically require ultrahigh vacuum and/or cryogenic environments, involve specialized sample preparation, and may suffer from electron-beam-induced damage, while often being incompatible with in situ or operando measurements. Consequently, in practical studies they are frequently complemented by techniques based on moiré periodicity or physical responses, together forming a comprehensive framework for twist-angle characterization.

### *2.2 Moiré-Based Characterization*

**The second category comprises methods based on the moiré period or morphology-related features**, including measurements of moiré superlattice periodicity using atomic force microscopy (AFM)[51], scanning electron microscopy (SEM)[52, 53], or optical microscopy[54]. Rather than directly resolving the crystallographic orientations of individual layers, these approaches infer the twist angle from the moiré superlattice period formed by interlayer twisting, or from morphology and property modulations induced by the moiré potential. By exploiting the well-defined geometric relationship between the moiré period and the twist angle, these methods exhibit exceptionally high sensitivity to angular variations in the small-angle limit, and are therefore widely adopted in low-twist-angle moiré systems.

Atomic force microscopy is among the most commonly employed techniques[59-61]. Although an ideal moiré superlattice is purely a geometric modulation, in realistic van der Waals heterostructures interlayer interactions give rise to local structural relaxation, in-plane strain, and out-of-plane corrugation, which generate resolvable periodic contrast in AFM topography or lateral force signals. Woods et al. directly observed moiré-periodic structures in graphene/hBN heterostructures using AFM imaging and quantitatively determined the lattice mismatch and relative rotational angle by comparison with a geometric model[51].

At larger length scales, optical microscopy and optical spectroscopic techniques can also be employed to indirectly determine the moiré period, particularly in small-twist-angle systems with long superlattice periods. When the moiré wavelength reaches tens to hundreds of nanometres, its modulation of the electronic band structure and Landau level spectrum can manifest in transport or optical responses. For example, Dean et al. inferred the moiré period and twist angle in graphene/hBN heterostructures from the Hofstadter butterfly spectrum observed in quantum Hall transport measurements[62]. Wang et al[54]. proposed a method based on optical microscopy, which visualizes the ordered n-triacontane (TTC) molecular crystals to determine the crystal orientation and interlayer twist angle of the two-dimensional crystals. When testing the two-dimensional double-layer structure composed of interleaved stacked $MoS_2$ crystals, this method can also accurately determine the twist angle.

Scanning electron microscopy (SEM) has likewise been used to characterize twist angles in two-dimensional materials. Unlike TEM-based approaches, SEM does not directly resolve atomic lattice diffraction patterns; instead, it relies on orientation-dependent contrast arising from the coupled scattering of the electron beam with the crystal lattice, enabling identification of distinct orientational domains and moiré structures. Tillotson et al. demonstrated that electron channeling contrast imaging (ECCI) in SEM produces pronounced contrast between regions with different twist configurations, allowing the spatial distribution and boundaries of twisted domains to be visualized[53].

Although SEM generally lacks the spatial resolution required to directly resolve atomic-scale moiré periodicity, it offers clear advantages in terms of large field of view and rapid localization of regions with distinct twist angles, and thus serves as a valuable auxiliary tool within multiscale characterization workflows.

Overall, methods based on moiré periodicity or morphology-related features are experimentally accessible and capable of providing large-area, spatially resolved information on twist angles, making them particularly well suited for small-angle systems and initial sample screening. Nevertheless, their sensitivity to interlayer relaxation, strain, and environmental conditions means that the extracted twist angles often need to be combined with atomic-resolution or diffraction-based techniques to achieve a more complete and cross-validated characterization of twisted van der Waals structures.

### *2.3 Spectroscopic and Electronic Characterization*

The third category comprises methods based on spectroscopic or electronic responses, including second-harmonic generation (SHG) spectroscopy and Raman spectroscopy[55-57]. In twisted bilayer or heterobilayer systems, each constituent layer generates its own SHG response, and the total SHG signal can be regarded as a coherent superposition of the nonlinear polarizations from the two layers[10, 13]. Because a finite twist angle $\theta$ exists between the crystallographic principal axes of the layers, the corresponding second-order nonlinear susceptibility tensors $\chi(2)$, are rotated relative to one another in the laboratory frame[2, 63]. This relative rotation leads to systematic variations in the angular dependence, polarization selection rules, and phase relationships of the resulting SHG intensity. By analysing the periodic modulation of SHG intensity as a function of incident or detected polarization angle, or by comparing SHG signals under parallel and cross-polarized configurations, the relative orientation of the two crystal axes, and hence the interlayer twist angle, can be directly extracted.

Raman spectroscopy characterizes material structure through the inelastic scattering of incident light by lattice vibrational (phonon) modes[49, 61, 64, 65]. Similar to SHG, the polarization dependence of Raman scattering encodes information about crystal symmetry. The Raman peak positions, intensities, and polarization selection rules are all strongly influenced by lattice symmetry and interlayer coupling. Kim et al. systematically investigated the evolution of Raman responses in twisted bilayer graphene as a function of twist angle, demonstrating that specific Raman features, such as enhancement of the G peak and mode splitting-serve as fingerprint signatures of the twist angle[66]. Subsequent studies have exploited low-frequency Raman modes to perform spatially resolved measurements in small-angle twisted systems, enabling mapping of the local twist angle[60, 66, 67].

More broadly, these spectroscopic approaches infer the twist angle by analysing how phononic, excitonic, or electronic band-structure responses are modulated by moiré periodicity. They offer key advantages, including non-contact operation and compatibility with in situ measurement conditions, making them particularly well suited for spatial mapping and investigations of dynamic processes.

For these reasons, it is essential to systematically examine twist-angle determination methods and their underlying physical meanings before discussing twist-angle control techniques. Reliable twist-angle characterization not only underpins the validation of fabrication accuracy and device reproducibility but also provides the foundation for benchmarking different control platforms, identifying the origins of experimental variability, and enabling feedback control and programmable twisting. This need is especially pronounced in dynamic twisting and in situ control platforms, where the twist angle is no longer a static parameter set once during fabrication, but a continuously monitored and actively regulated variable, further underscoring the central role of twist-angle metrology in moiré engineering.

### 3. Twist-Angle Engineering and Control in Two-Dimensional Materials

The realization of twist-engineered van der Waals (vdW) heterostructures relies fundamentally on the ability to precisely generate, control, and preserve the interlayer rotational configuration throughout fabrication and device operation[68]. As the twist angle has evolved from a structural descriptor into a programmable degree of freedom governing electronic, optical, and quantum functionalities, the development of reliable twist-control strategies has become a central challenge in both twistronics and moiré photonics[61, 64, 65, 69]. Beyond achieving a target rotational alignment, practical approaches must simultaneously minimize contamination, suppress strain accumulation, and maintain interfacial integrity, while providing sufficient precision and reproducibility for systematic investigations of twist-dependent phenomena. The technological evolution of reconfigurable platforms in twisted two-dimensional materials and photonic devices is shown in (Fig 1).

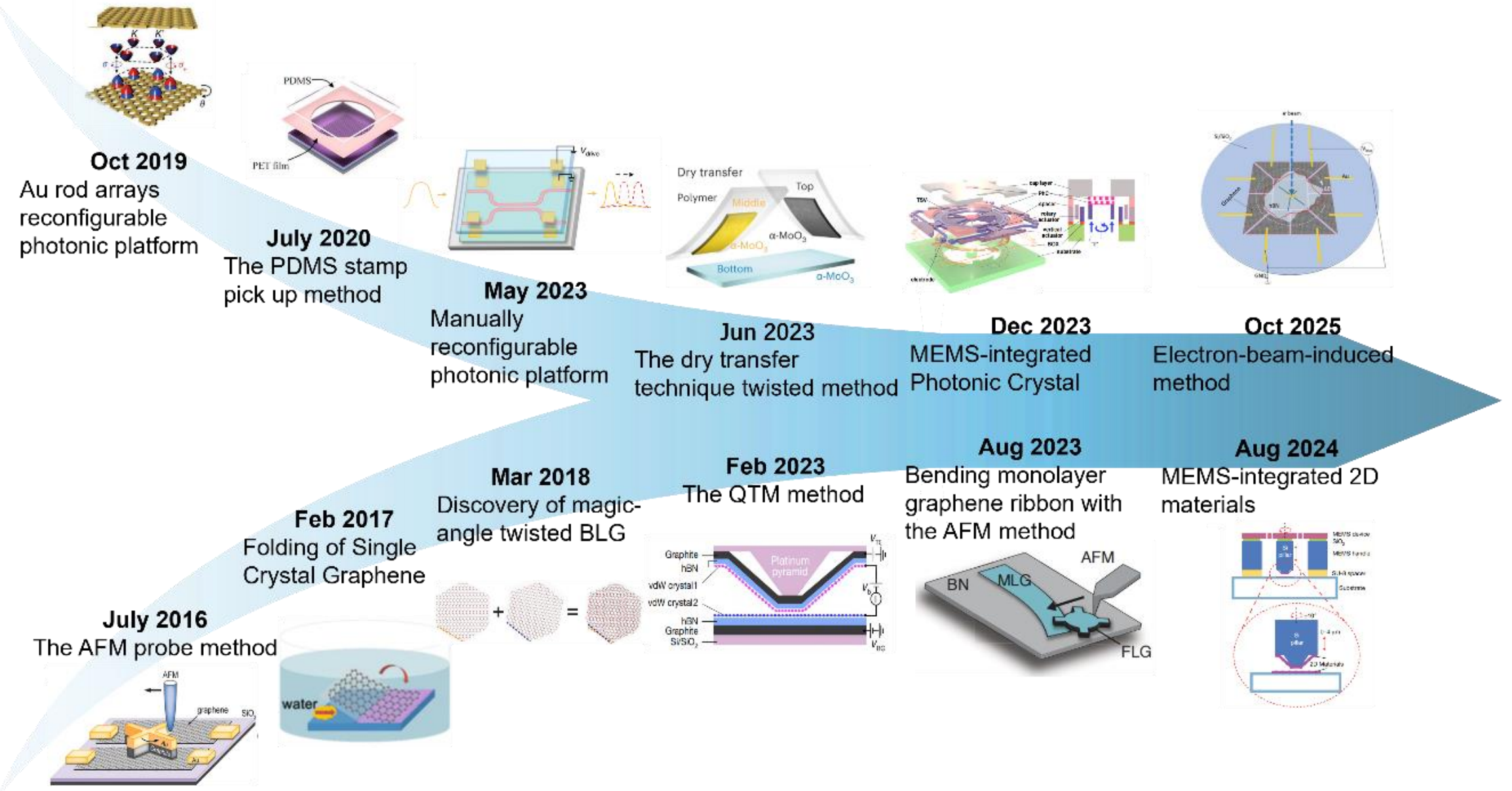

**Fig. 1.** Technological evolution of reconfigurable platforms for twisted two-dimensional materials and photonic devices. Representative milestones from 2016 to 2025 illustrating the parallel development of reconfigurable 2D-material assembly (bottom) and programmable photonic platforms (top)[70-81]. The timeline highlights the evolution from deterministic transfer and manual assembly to dry-transfer, AFM- and MEMS-assisted manipulation, and electron-beam-induced reconfiguration, reflecting the ongoing transition toward dynamically reconfigurable, high-precision platforms for twist-angle engineering and adaptive photonic functionalities.

Different twist-engineering strategies inevitably involve trade-offs between angular precision, tunable range, addressable area, repeatability, scalability, and interfacial cleanliness. These considerations not only determine the experimental accessibility of angle-dependent moiré states but also dictate the feasibility of translating laboratory-scale demonstrations into programmable photonic devices and large-scale manufacturing. Accordingly, the evolution of twist engineering has progressed from static fabrication methods that define the twist angle during assembly to dynamically reconfigurable platforms capable of post-fabrication manipulation. The representative strategies for the preparation and reconstruction of 2D materials are shown in (Fig 2). In the following sections, we review the major technological routes for twist control, including deterministic transfer and growth-based approaches, AFM-assisted manipulation, quantum twisting microscopy (QTM), microelectromechanical systems (MEMS), and emerging contactless strategies, highlighting their respective advantages, limitations, and future development directions toward programmable and scalable moiré photonic platforms.

### *3.1 Static Twist-Angle Engineering*

In the early stages of research on the physical properties of twisted two-dimensional materials, experimental constraints generally limited studies to static structures with fixed twist angles[65, 68, 82]. Representative approaches at this stage primarily included folding-based methods and exfoliation–stacking techniques. However, twisted bilayer structures fabricated using these methods—such as twisted bilayer graphene (BLG)—typically exhibited a random distribution of twist angles. As a result, extensive sample fabrication and screening were often required to fortuitously identify devices with a desired target angle. More critically, even sub-degree variations (<1°) in the twist angle are sufficient to induce qualitative changes in the electronic band structure, giving rise to entirely new physical phenomena[61, 69, 83]. This lack of deterministic control over the twist angle not only substantially increased experimental cost and complexity, but also severely constrained the systematic investigation and practical application of twisted two-dimensional materials[64, 68, 83, 84].

To achieve deterministic definition and precise control of the twist angle, significant efforts have since been devoted to refining early fabrication strategies through advances in both processing techniques and device platforms. In the following subsections, we review several representative static fabrication methods for fixed-angle twisted structures, with emphasis on their underlying workflows as well as their respective strengths and limitations in terms of angular precision, repeatability, and applicable parameter space[2, 64, 68].

**A. Folding-induced twist approaches*:*** Folding provides a comparatively simple route to twisted bilayers without executing a full tear-and-stack workflow. Early demonstrations relied on gently rinsing a substrate bearing exfoliated graphene with deionized water, which could trigger spontaneous folding and thereby yield twisted bilayer graphene[85, 86]. However, interfaces produced in this manner are often insufficiently clean, and the rinsing step may promote unintended adhesion between the folded layers, complicating subsequent measurements.

To improve controllability and sample quality, several refinements and alternatives have been developed. One approach combines exfoliation with thermal processing: graphene flakes are deposited onto a preheated poly(methyl methacrylate) (PMMA 950K) film, followed by annealing to generate folded graphene[87]; alternatively, flakes are exfoliated onto $Si/SiO_2$ substrates (typically with a 300-nm thermally grown oxide) and heated on an open hot plate to ~150 °C, producing "self-folded" structures through a sequence of spontaneous folding, sliding, delamination, and tearing[88]. Beyond thermal routes, chemically induced folding has been reported by suspending a graphene-on-Si substrate above the surface of a freshly prepared $H_2SO_4$:$HNO_3$ (3:1, v/v) mixture, where exposure to the acidic vapor induces folding or wrinkling[89]. Instrument-assisted strategies have also been employed, including mechanical manipulation with an atomic force microscope or fortuitous layer flipping during exfoliation from natural graphite[90].

More recently, folding concepts have been generalized to other two-dimensional materials and implemented in more controlled mechanical settings. In representative experiments on $MoS_2$, mechanically exfoliated flakes are transferred onto an elastomeric substrate, subjected to a large tensile strain, and then released to trigger wrinkle formation and partial delamination[91]. The resulting folded regions naturally introduce a relative rotational misalignment, enabling optical and spectroscopic characterization under twisted conditions to probe interlayer coupling.

Despite these advances, folding-based approaches share a common limitation: the resulting twist angles are typically distributed and difficult to predefine, which hinders systematic studies that require deterministic angle selection[92].

With the advent of large-area single-crystal monolayer graphene grown by chemical vapor deposition (CVD), a more scalable and uniform folding strategy has emerged[93]. In this scheme, macroscopic single-crystal monolayer graphene is grown on Cu(111) and transferred onto a custom substrate patterned with hydrophobic and hydrophilic regions, enabling a programmed "self-folding" process. The mechanism exploits the preferential delamination of the film from hydrophilic regions upon immersion in water; withdrawing the substrate yields millimeter-scale folded graphene with improved spatial uniformity. Notably, this method can produce multiple FG samples with specified twist angles, and-relative to tear-and-stack-CVD provides monolayers of larger area and higher uniformity, thereby improving the scalability of folded twisted structures.

**B. Tear and stack approaches:** Compared with folding, the tear-and-stack workflow has become the dominant route to fabricating twisted bilayer 2D materials because it offers more deterministic control over the twist angle and thus enables higher angular precision. Early demonstrations combined mechanical exfoliation with a double-sacrificial-layer transfer scheme to assemble graphenehexagonal boron nitride (hBN) stacks, achieving an angular accuracy of approximately ~0.5°[94]. Interfaces produced by this approach could reach near-atomic cleanliness; however, exfoliated flakes often suffer from limited spatial uniformity and small usable overlap areas, and the double-sacrificial-layer process is comparatively complex and labor-intensive.

Subsequently, a dry-transfer strategy employing a hemispherical "handle" substrate-designed to deliberately minimize the pick-up contact area-was introduced[95]. By coating the apex of an epoxy hemispherical tip with a poly(vinyl alcohol) (PVA) layer, micron-scale flakes can be selectively retrieved from an exfoliation substrate while preserving clear crystallographic orientation information. After picking up the first layer, the bottom substrate is rotated to set the target twist angle, and the second layer is then collected to complete the stack. While this geometry-constrained pick-up reduces operational complexity, the procedure is not readily repeatable, the achievable angular precision remains limited, and the resulting structures still typically exhibit small area and modest uniformity.

The development of CVD-capable of producing large-area, highly uniform monolayers with well-defined triangular or truncated-triangular crystal morphologies-motivated a CVD + stacking route. In a representative protocol, monolayer $MoS_2$ is grown by CVD, divided into two pieces, and assembled via dry transfer: after spin-coating poly(methyl methacrylate) (PMMA), one piece is mounted on a polydimethylsiloxane (PDMS) elastomer supported by a glass slide for manipulation, enabling the two $MoS_2$ layers to be stacked at different twist angles[96]. A notable limitation is that the twist angle is typically determined by electron-microscopy inspection of the relative orientation of triangular domains, leading to an angular accuracy on the order of ~2°, which is inadequate for small-angle twistronics.

For stacking large-area CVD-grown monolayers, wet transfer can offer practical advantages over dry transfer. For example, in constructing $WS_2$-$WSe_2$ heterostructures, a PMMA film in a KOH etchant is used to delaminate triangular monolayer $WS_2$ from the growth substrate. The $WS_2$/PMMA film is then laminated onto a target substrate (e.g., $SiO_2$/Si or sapphire) hosting a high density of $WSe_2$ monolayers. After transfer, the sample is immersed in acetone to remove PMMA and residual contaminants, followed by annealing to enhance interlayer coupling[97, 98]. Because rotational control in wet transfer is generally poorer than in dry transfer, femtosecond-laser cutting has been adopted to define fiducial cut lines; the relative angle between these cut lines provides a more reliable handle for setting the twist angle and can improve angular accuracy. Nevertheless, wet transfer remains susceptible to PMMA residues, and interlayer wrinkling can arise from trapped moisture and its subsequent evaporation.

More recently, refinements to conventional dry-transfer methods have been developed to better accommodate large-area CVD materials, enabling tear-and-stack assembly with improved area, uniformity, interfacial cleanliness, and controllability[99-102]. For instance, thin-film-assisted transfer (TAT) has been reported to enable deterministic fabrication of vdW structures with a defined twist angle, with an accuracy of ~0.37°. In addition, flexible silicon nitride film assembly replaces polymer supports with an inorganic, compliant SiN membrane, facilitating scalable transfer of CVD-grown materials and markedly increasing the characteristic size of atomically clean interfacial regions.

Collectively, these developments highlight an important principle in twist fabrication: constraining contact geometry during pick-up can enhance repeatability and suppress unintended shear. In practice, however, early implementations still faced a throughput-selectivity trade-off: when the target twist-angle window is narrow, extensive fabrication and screening are often required, because even sub-degree deviations can qualitatively alter electronic and optical behavior.

**C. Reusable fixed-angle photonic platforms and STM-controlled folding:** Beyond electronic 2D materials, twist platforms have been adapted for photonic structures[103-105]. For example, bilayer arrays of patterned metallic or dielectric elements can be assembled at prescribed twist angles to realize tunable chiral and nonlinear optical responses. In parallel, scanning tunneling microscope control of folding direction has enabled atomically precise “origami” graphene nanostructures, offering high geometric control and the possibility of reversible fold/unfold operations. These methods highlight a broader theme: engineering the interface geometry can be as important as material choice in determining twist-enabled functionality.

**D. Bottom-up synthesis of twisted vdW architectures:** While mechanical stacking dominates current practice, bottom-up synthesis has emerged to address interfacial contamination and limited large-area uniformity. Representative routes include self-organized twist heterostructures formed through aligned vdW epitaxy followed by solid-state transformations, as well as helical or chiral growth modes driven by screw dislocations and curved or non-Euclidean substrates[106-111]. In Eshelby-twisted nanowires, the twist rate scales inversely with cross-sectional area, enabling tunable twist in quasi-1D geometries but making per-layer twist intrinsically small for large-area 2D sheets.

Bottom-up routes thus offer a promising direction for scalable twist engineering, yet they also introduce new constraints: the accessible angle range may be set by growth energetics, byproduct domains with uncontrolled twist can appear, and decoupling twist from concurrent strain or compositional gradients remains nontrivial. Continued progress will likely rely on combining growth design with post-growth characterization and selective transfer to isolate the most uniform regions.

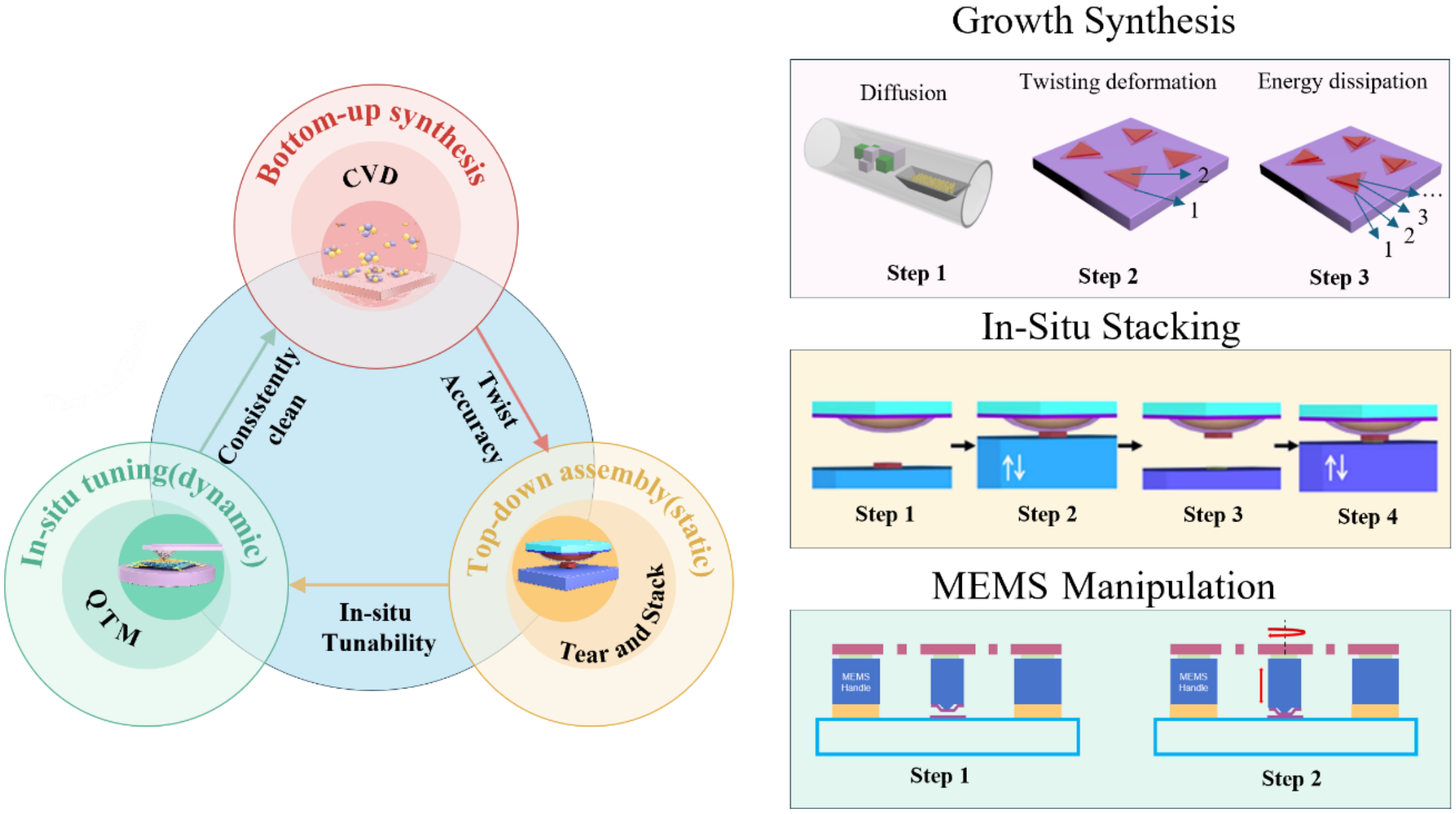

**Fig. 2.** Comparison of representative strategies for fabricating and reconfiguring twisted two-dimensional (2D) materials. Left, schematic comparison of bottom-up synthesis, top-down assembly, and in situ tuning strategies. Right, representative fabrication and operation workflows for bottom-up synthesis (top), top-down tear-and-stack assembly (middle), and MEMS-enabled in situ tuning (bottom).

### *3.2 Dynamic Platforms for In Situ Twist Control*

Although static "tear-and-stack" workflows have continued to mature—for example, the use of scanning tunnelling microscopy to steer folding directions and realize small-area origami-like graphene architectures—fixed-angle structures have benefited from improved geometric definition and, in selected cases, the prospect of reversible folding/unfolding. Nevertheless, the fundamental limitation of tear-and-stack remains: once the twist angle is defined, it cannot be tuned continuously. Systematic studies of twist-dependent phenomena therefore often require fabricating and characterizing large numbers of devices at different angles, while cross-device comparisons are inevitably confounded by sample-to-sample variations. To partially mitigate this bottleneck, thermally induced rotation has been explored, leveraging high-temperature annealing to activate low-friction sliding and rotation of 2D flakes on atomically flat substrates; however, its angular selectivity is typically poor. The field has thus increasingly shifted toward in situ reconfigurable platforms that convert twist control from discrete setting to continuous and programmable tuning. In particular, AFM-assisted manipulation schemes and micro-electro-mechanical system (MEMS)-enabled twisting platforms represent a transition from manually executed, largely static operations to device-driven, software-addressable dynamic modulation. This shift improves both flexibility and repeatability, and provides a foundation for future high-precision, automated, and scalable twist-control technologies.

**A. Manually reconfigurable photonic platforms:** For twisted photonic crystals, manual rotation remains a practical and widely used route to rapidly survey angle-dependent resonances. Representative examples include bilayer alumina-rod photonic crystals in which the relative rotation can be adjusted over a broad angular range, as well as bilayer moiré photonic lattices mounted on rotatable holders, enabling real-time tuning with optical readout of moiré patterns[112, 113]. Such platforms are well suited for rapid prototyping and proof-of-concept demonstrations; however, their achievable angular resolution and long-term stability are generally inferior to closed-loop actuation systems.

**B. AFM-driven actuation:** In situ twist control is particularly valuable for mapping observables as a continuous function of twist angle within a single device. AFM-based architectures have demonstrated this capability in graphene-hBN heterostructures, where superlubricity at the graphene/hBN interface helps overcome interfacial friction. Taking advantage of the high stiffness of hBN, a calibrated force can be applied to a micron-thick hBN "gear" or cantilever to generate an in-plane torque that drives rotation atop monolayer graphene, enabling twist-angle control; analogous principles have subsequently been extended to bilayer graphene homostructures[114-116]. AFM-tip approaches enable continuous twist tuning over a large angular span and can operate on comparatively large samples; reported twist-angle control can reach ~0.1°[117]. However, the mechanism is sensitive to nanoscale contact geometry: even small lateral displacements (<100 nm) can misregister the contact region or introduce unintended in-plane stress in the absence of dedicated stress-free protocols. Consequently, large interfacial displacements may be difficult to predict a priori, which can compromise repeatability and complicate quantitative interpretation.

**C. Patch-assisted polymer-handle manipulation:** Beyond AFM probes, an alternative in situ manipulation strategy patterns a PMMA polymer patch on a target flake via electron-beam lithography (EBL) and uses a simple hemispherical PDMS "handle" to contact the patch[118]. When the interface remains incommensurate, kinetic friction is strongly suppressed (superlubricity), allowing the upper stack (for example, top hBN together with the underlying graphene) to slide and rotate relative to a bottom hBN support during manipulation, thereby enabling dynamic control of both rotation and position. A key practical advantage is operational convenience and reusability: the PMMA patch can be removed with acetone and redefined by EBL for subsequent tuning cycles. The approach is compatible with flakes of arbitrary thickness and is, in principle, transferable across diverse 2D material systems, whereas direct AFM-tip manipulation can damage ultrathin layers. The principal constraints include a relatively small addressable twisting area, loss of reversible tuning near commensurate self-locking, increased process complexity, and the possibility of polymer-related interfacial contamination.

**D. Quantum Twisting Microscope (QTM):** The Quantum Twisting Microscope (QTM) represents a major advance in high-precision in situ twisting[79, 119, 120]. QTM is a scanning-probe platform that elevates the AFM cantilever into an integral part of the twist device: a Pt pyramid is fabricated near the cantilever edge using EBL-assisted metal deposition and focused-ion-beam (FIB) deposition, and a vdW stack (graphite/hBN/multilayer graphene, MLG) is assembled on the pyramid to form a tip-integrated vdW device. The tip can establish a clean 2D interface by contacting a conventional vdW material stack on a planar substrate. Twist tuning is achieved by rotating the bottom sample with a piezoelectric rotator, enabling continuous control over a full 360° range with an angular resolution of ~0.001°. Importantly, the QTM tip can also function as a minimally invasive, momentum-resolved probe, enabling spatially resolved (~100 nm) scanning measurements that directly access energy–momentum dispersion. QTM mitigates uncontrolled stress that can arise in conventional AFM-based twisting and is broadly compatible with layered conductors, semiconductors, and superconductors. The main trade-off is the small twisting contact region, with a maximum linear dimension on the order of ~1 μm.

**E. Strain-field engineering to mitigate moiré disorder:** Twist-angle non-uniformity and moiré unit-cell distortions driven by uncontrolled strain fluctuations have long hindered quantitative comparisons between experiment and theory. A complementary route to explicit rotation is global strain-field engineering. In one implementation, an hBN substrate is used to sequentially pick up a straight graphene ribbon, after which an AFM tip slides a gear-like few-layer graphite micro-arm at the ribbon end to bend the ribbon in-plane[121]. Upon load release, static friction can lock in the deflection profile, thereby imposing an in-plane load that generates a continuously varying twist angle along a predefined direction. Additional layers can subsequently be integrated using conventional dry-transfer workflows without adversely altering the programmed bending geometry. This strategy enables smooth and continuous twist tuning with a reported intrinsic twist-angle disorder of 0.0074°, substantially lower than that of conventional tear-and-stack twisted bilayer graphene. It is applicable across combinations of graphene, hBN, and transition-metal dichalcogenides. The improvement in uniformity and reduction of random strain and moiré disorder sharpen band-structure inferences; however, the method inherently co-tunes strain together with twist, and the accessible twist-angle span is limited to ~5°-beyond which highly strained wrinkles can form.

**F. MEMS-enabled universal interface actuation (MEGA2D):** Chip-scale MEMS platforms provide a route to integrated, multi-degree-of-freedom control of vdW interfaces. In MEGA2D, electrostatic actuation enables stepwise twist control with a reported ~0.33° step size, together with sub-nanometre planarity control, aiming to deliver an on-chip, reconfigurable vdW interface[122]. Beyond 2D electronic stacks, related implementations target twisted moiré photonic-crystal sensors, where both twist angle and interlayer spacing can be tuned to precisely program optical responses[123]. More broadly, the platform is designed to generalize across 2D materials, allowing controlled rotation, compression, and lateral displacement in bilayers for real-time modulation of interfacial properties. While interfacial cleanliness remains constrained by dry-transfer residues, some bubbles and wrinkles can gradually diminish during contact-and-rotate operations, increasing the effective clean overlap area. Key advantages include a wide tunable range (±10°), improved operability, enhanced repeatability and scalability potential, compatibility with cryogenic environments, and a plausible pathway toward large-scale integration.

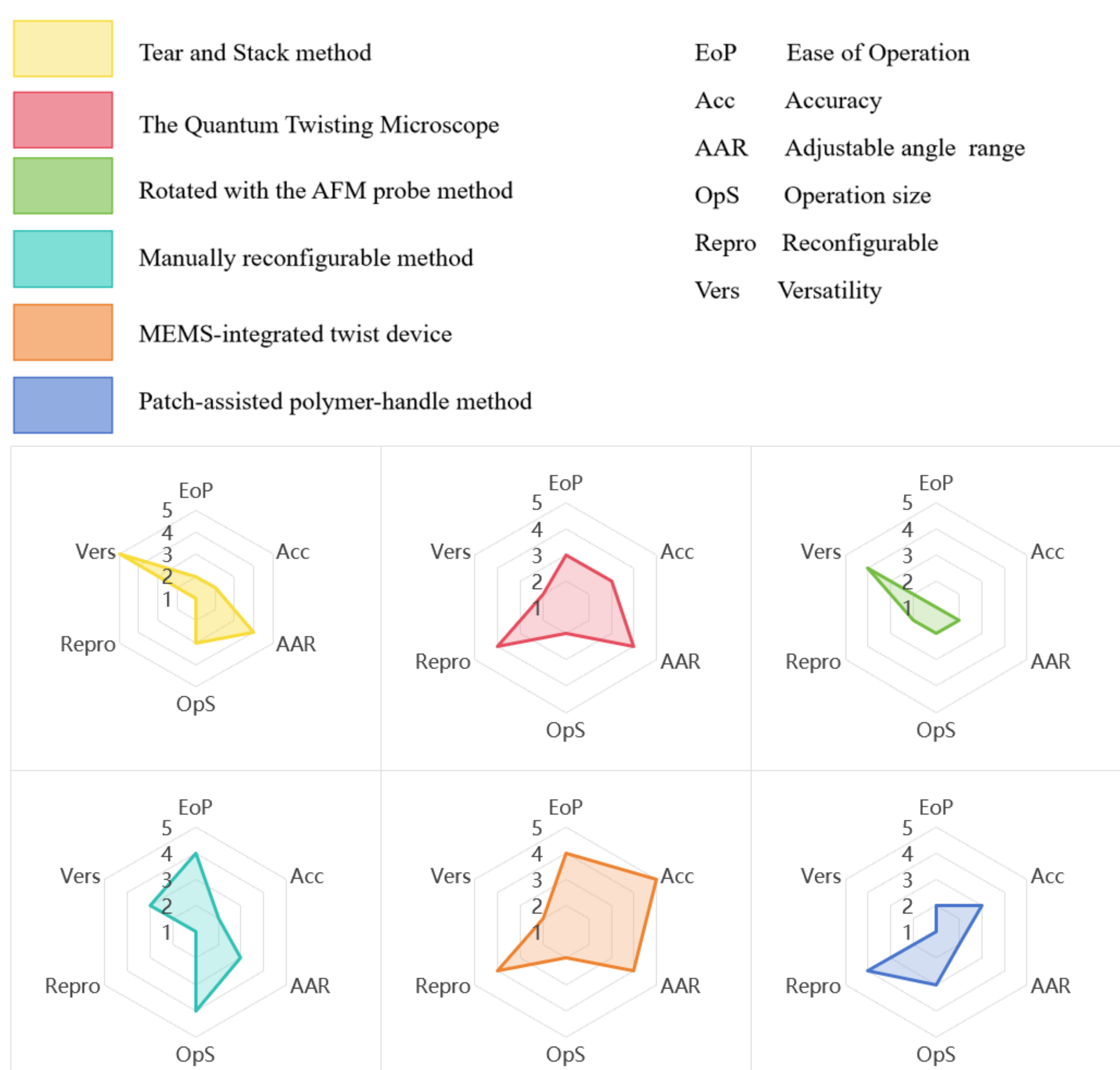

**Fig 3** Radar-chart comparison of representative twist-angle engineering methods for two-dimensional (2D) materials. Six representative approaches are evaluated, including the tear-and-stack method[94], the Quantum Twisting Microscope (QTM)[79], AFM probe-assisted rotation[115], the manually reconfigurable method[124], the MEMS-integrated twist device[81], and the patch-assisted polymer-handle method[125]. The comparison is based on six performance metrics: ease of operation (EoP), accuracy (Acc), adjustable angle range (AAR), operation size (OpS), reconfigurability (Repro), and versatility (Vers), with scores ranging from 1 (lowest) to 5 (highest).

*3.3 Contactless Actuation for Twist Control*

**Electron-beam charging and optical torque:** Beyond mechanical actuation, contactless approaches offer a distinct route to twist control. Electron-beam-induced charge injection can locally charge an insulating hBN flake on a grounded conductor, producing an interfacial electric field that generates in-plane electrostatic torque and rotational displacement[126]. Optical schemes have also been proposed, leveraging vortex beams carrying orbital angular momentum (OAM) to apply a tunable optical torque that could, in principle, enable rapid and scalable control without physical contact. The central challenge for these emerging strategies is to achieve deterministic, quantifiable angle control while managing heating, charging stability, and material-specific dissipation channels.

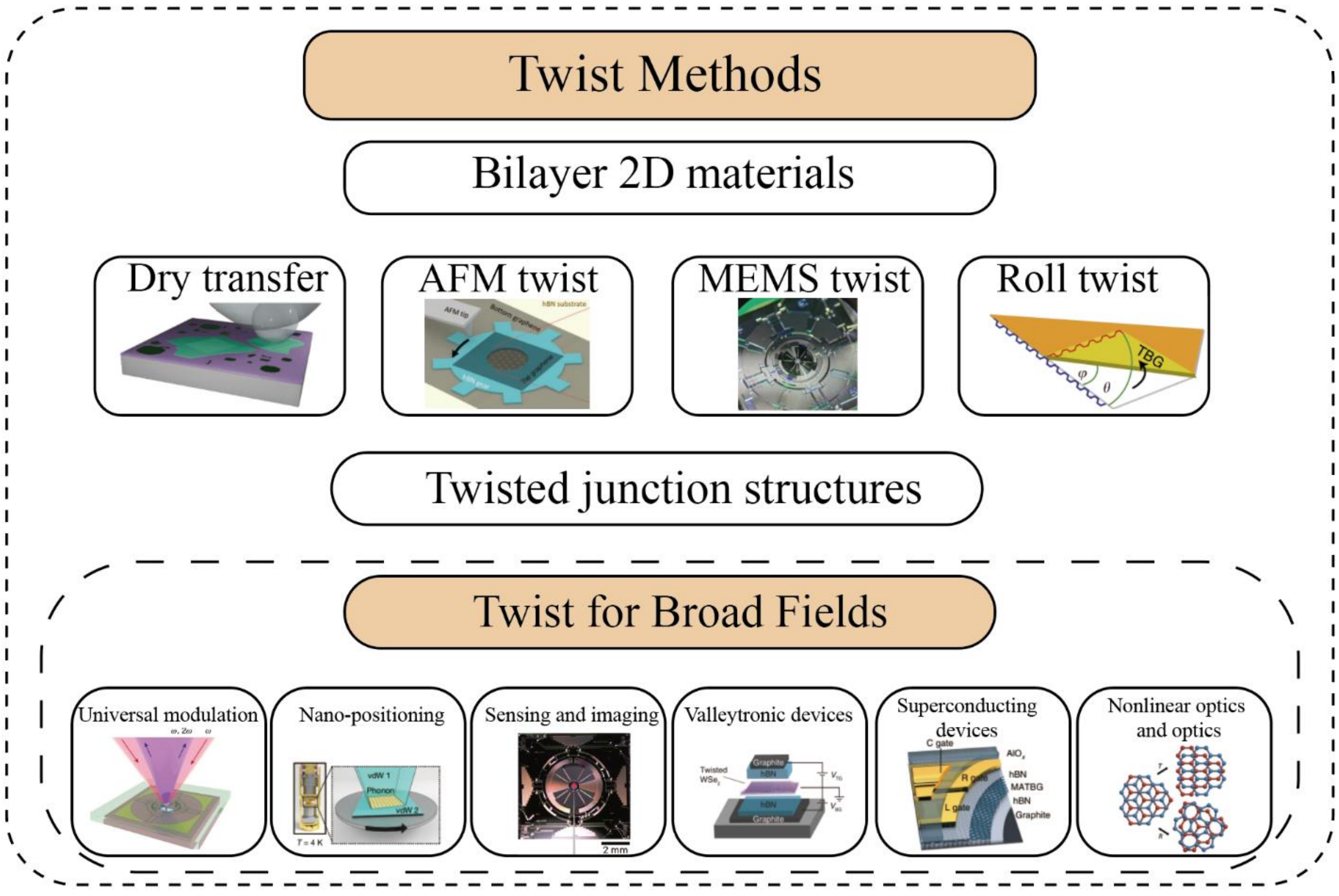

**Fig. 4.** Representative twist methods and their applications. The upper panel illustrates four representative approaches for constructing twisted bilayer 2D materials-dry transfer, AFM-assisted twisting, MEMS-based twisting, and roll-based twisting-to realize twisted junction structures[81, 90, 95, 117]. The lower panel summarizes representative applications of twisted structures in universal modulation, nano-

positioning, sensing and imaging, valleytronics, superconducting devices, and nonlinear optics and optoelectronics[8, 74, 81, 127-129].

In summary, twisting devices have created powerful new opportunities for the exploration and active manipulation of two-dimensional materials and photonic crystals, with implications spanning universal modulation, nanoscale positioning, sensing, valleytronic functionalities, superconductivity, and nonlinear optics. Among the diverse strategies discussed above, MEMS-enabled twist platforms appear especially promising, as they move twist engineering beyond static sample preparation toward a programmable, scalable, and genuinely device-oriented technology. Compared with conventional fixed-angle assembly approaches, MEMS-based systems offer several decisive advantages, including on-chip integration, multidimensional actuation, enhanced repeatability, wider tunable angular range, compatibility with cryogenic operation and semiconductor fabrication processes, and clear prospects for large-scale integration. More fundamentally, these platforms transform twist angle and interlayer spacing from one-time fabrication parameters into continuously addressable engineering variables[63, 130, 131]. This capability provides a practical route toward reproducible studies of moiré physics, as well as reconfigurable photonic and optoelectronic devices whose properties can be tuned on demand. For these reasons, MEMS-based twist control is compelling not only from a scientific perspective, but also as a technological foundation for future twisted-material systems. By establishing a direct pathway from fundamental twist-dependent phenomena to electrically tunable and experimentally realizable device architectures, continued advances in this area are likely to further strengthen the role of MEMS at the interface of quantum materials, advanced optics, and integrated photonics, ultimately accelerating the translation of twisted-bilayer concepts into robust and application-ready technologies[82].

## 4. Toward Programmable Twist-Engineered Photonic Systems

In recent years, twist-induced moiré phenomena have evolved beyond their original context in two-dimensional electronic materials and have increasingly emerged in photonic systems, polaritonic platforms, and artificially engineered nanophotonic architectures. Previous studies have demonstrated that photonic moiré structures can support a wide range of unconventional wave phenomena, including localized-to-delocalized phase transitions, flat-band formation, and photonic "magic-angle" effects, highlighting that moiré physics is not exclusive to electronic systems but also represents a fundamental paradigm for manipulating light–matter interactions.

From a photonic perspective, the introduction of twist as an additional structural degree of freedom fundamentally expands the design space of multilayer optical architectures. Without altering the constituent materials or the geometry of individual layers, continuously varying the interlayer twist angle enables deterministic reconfiguration of photonic band dispersion, interlayer mode hybridization, symmetry-protected resonances, and far-field radiation characteristics, thereby providing an unprecedented route toward programmable and reconfigurable photonic devices.

Despite these remarkable advances, current research on twisted bilayer photonic crystal membranes remains largely focused on static device configurations with predefined twist angles. Such approaches have been instrumental in uncovering fundamental moiré photonic phenomena but are inherently limited in their ability to support adaptive functionality, real-time optimization, and scalable device manufacturing. Looking forward, the next stage of development is expected to shift from the realization of individual twisted structures toward the establishment of dynamically reconfigurable, feedback-controlled, and manufacturable photonic platforms. In our view, this transition will be driven by four closely connected technological directions that collectively define the future roadmap of twisted bilayer photonic crystal membranes: (i) atomic-force-microscopy (AFM)-based in situ twist manipulation, enabling precise exploration of angle-dependent moiré photonic phenomena; (ii) microelectromechanical-system (MEMS)-enabled programmable photonic platforms, providing integrated, reversible, and electrically addressable control of the twist degree of freedom; (iii) quantum twisting microscopy (QTM), integrating ultrahigh-precision twist control with simultaneous local characterization to establish closed-loop twist engineering; and (iv) roll-to-roll (R2R) wafer-scale manufacturing, combined with machine vision, artificial intelligence, and automated feedback control, to achieve scalable, high-throughput fabrication of twisted photonic architectures. Together, these developments outline a clear evolution from proof-of-concept demonstrations toward programmable, intelligent, and industrially scalable moiré photonic systems.

### *4.1 From Static Structures to In Situ Twist Manipulation*

Despite these pioneering demonstrations((Fig. 5a)[117] (Fig. 5b)[132]), AFM-based nanomanipulation remains fundamentally limited by its serial and manually operated nature. Twist manipulation typically relies on direct tip–sample interaction and therefore suffers from limited throughput, operator dependence, and poor compatibility with automation. Moreover, the inherently localized operation of AFM makes it unsuitable for wafer-scale integration or high-volume manufacturing, while its mechanical manipulation strategy offers limited potential for programmable and continuously addressable device operation. These limitations suggest that AFM is unlikely to evolve into a practical platform for large-scale reconfigurable photonic technologies, motivating the development of electrically actuated and highly integrated platforms such as microelectromechanical systems (MEMS), which will be discussed in the following section.

Nevertheless, the long-term significance of AFM should not be evaluated solely from the perspective of manufacturability. Instead, its greatest value lies in its unparalleled capability for fundamental exploration of twist-dependent moiré phenomena. Unlike integrated actuation platforms that are generally designed to operate within predefined parameter windows, AFM offers exceptional flexibility for continuously scanning the entire twist-angle phase space while simultaneously probing the accompanying evolution of interlayer relaxation, friction, strain redistribution, and optical responses. Such capabilities are particularly important for discovering previously unknown photonic magic angles, identifying emergent phase transitions, and establishing quantitative correlations between local stacking configurations and optical functionalities. In this sense, AFM is expected to remain an indispensable experimental platform for revealing new moiré photonic physics, whereas MEMS and other integrated platforms will primarily translate these discoveries into programmable device architectures.

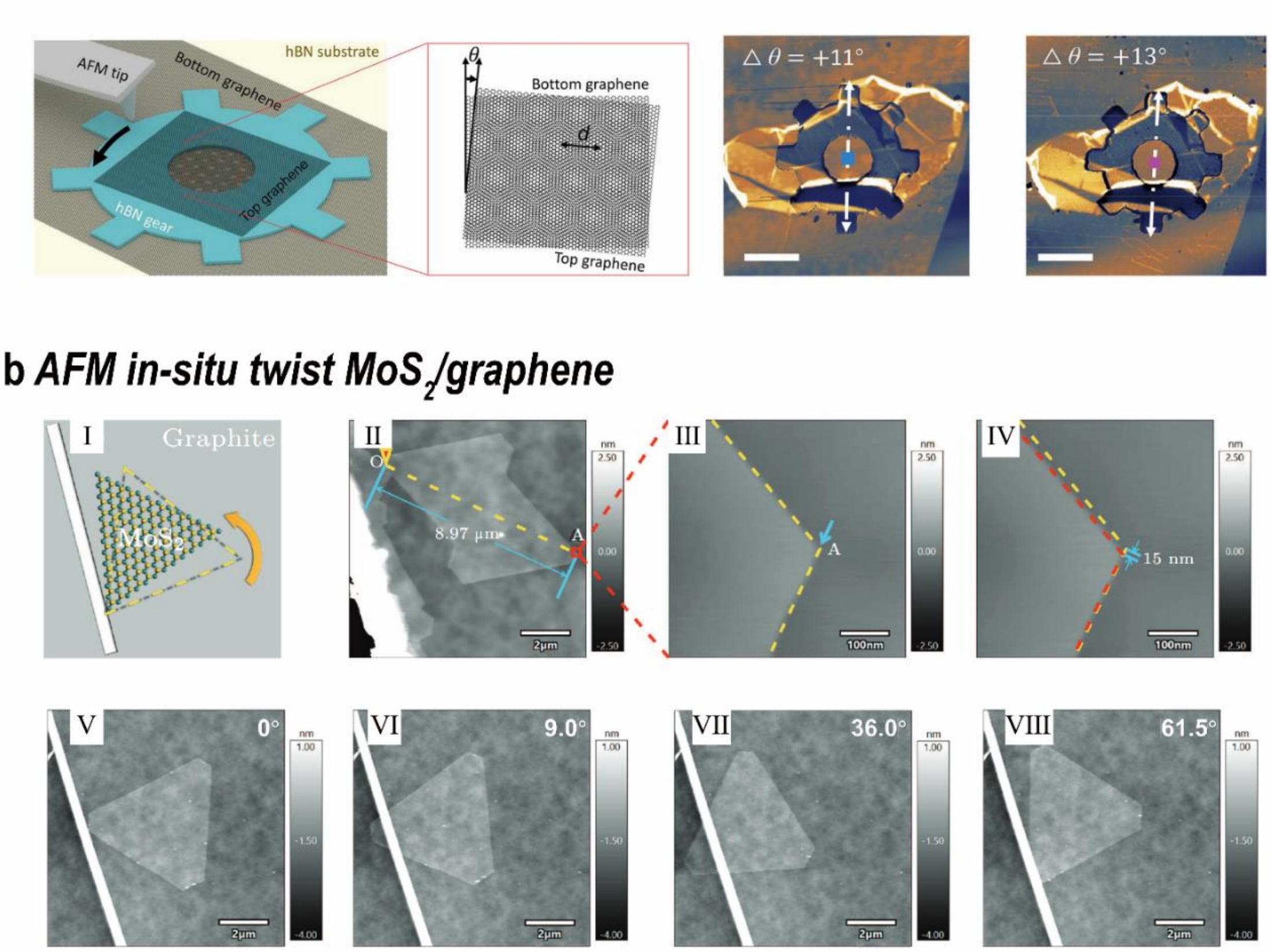

**Fig. 5.** The AFM-based in-situ twist protocol. 2D materials are continuously pushed through AFM. **a** I-III. The twisting heterojunctions[117]. **b** Twist angle control of $MoS_2$/graphene heterostructure[132]. I Schematic diagram of the method to control the twist angle between $MoS_2$ and graphene by an AFM tip. II-IV. A demonstration of precise control of interlayer twist angle below 0.1◦. V-VIII. AFM images of $MoS_2$/graphene heterostructure at different twist angles we want.

Looking forward, AFM-based twist engineering is likely to evolve from a purely mechanical manipulation tool into a multimodal characterization platform integrating in situ optical spectroscopy, near-field imaging, and intelligent feedback control. The combination of AFM with techniques such as Raman spectroscopy, second-harmonic generation, photoluminescence mapping, or scanning near-field optical microscopy could enable simultaneous manipulation and real-time characterization of angle-dependent photonic states. Coupled with machine vision and artificial intelligence, future AFM platforms may further realize autonomous twist exploration by continuously searching for optimal stacking configurations and emergent optical phenomena. Such developments would transform AFM from a proof-of-concept manipulation technique into a powerful discovery engine for next-generation moiré photonics.

### *4.2 Toward Programmable Photonic Devices*

Although current MEMS platforms have already demonstrated state-of-the-art capabilities for sub-nanometre vertical positioning and high-precision control of the twist degree of freedom (Fig. 6a)[81] (Fig. 6b)[74]), their long-term significance extends far beyond incremental improvements in actuation accuracy. The next generation of MEMS is expected to evolve from a sophisticated experimental platform into a universal infrastructure for programmable moiré photonics. Rather than pursuing increasingly higher mechanical precision alone, future developments should emphasize accessibility, standardization, and system-level integration, enabling twist engineering to become a broadly deployable experimental capability rather than a laboratory-specific technique. In particular, modular and plug-and-play MEMS architectures with standardized electrical interfaces and compatible sample holders could substantially lower the barrier to adoption, allowing researchers without specialized microfabrication expertise to perform dynamic twist manipulation on demand.

Beyond improving usability, future MEMS platforms should expand from single-parameter actuation toward multidimensional control of the complete interlayer configuration. Independent and simultaneous tuning of twist angle, interlayer spacing, lateral translation, shear, strain, and external stimuli will enable systematic exploration of the multidimensional moiré energy landscape, providing unprecedented opportunities for discovering emergent photonic states that cannot be accessed by varying the twist angle alone. Equally important will be the integration of MEMS with in situ optical characterization and intelligent feedback algorithms, establishing closed-loop systems capable of autonomous calibration, real-time structural compensation, and adaptive optimization during device operation. Such self-correcting architectures will be particularly important for maintaining long-term stability in practical photonic devices operating under varying environmental conditions.

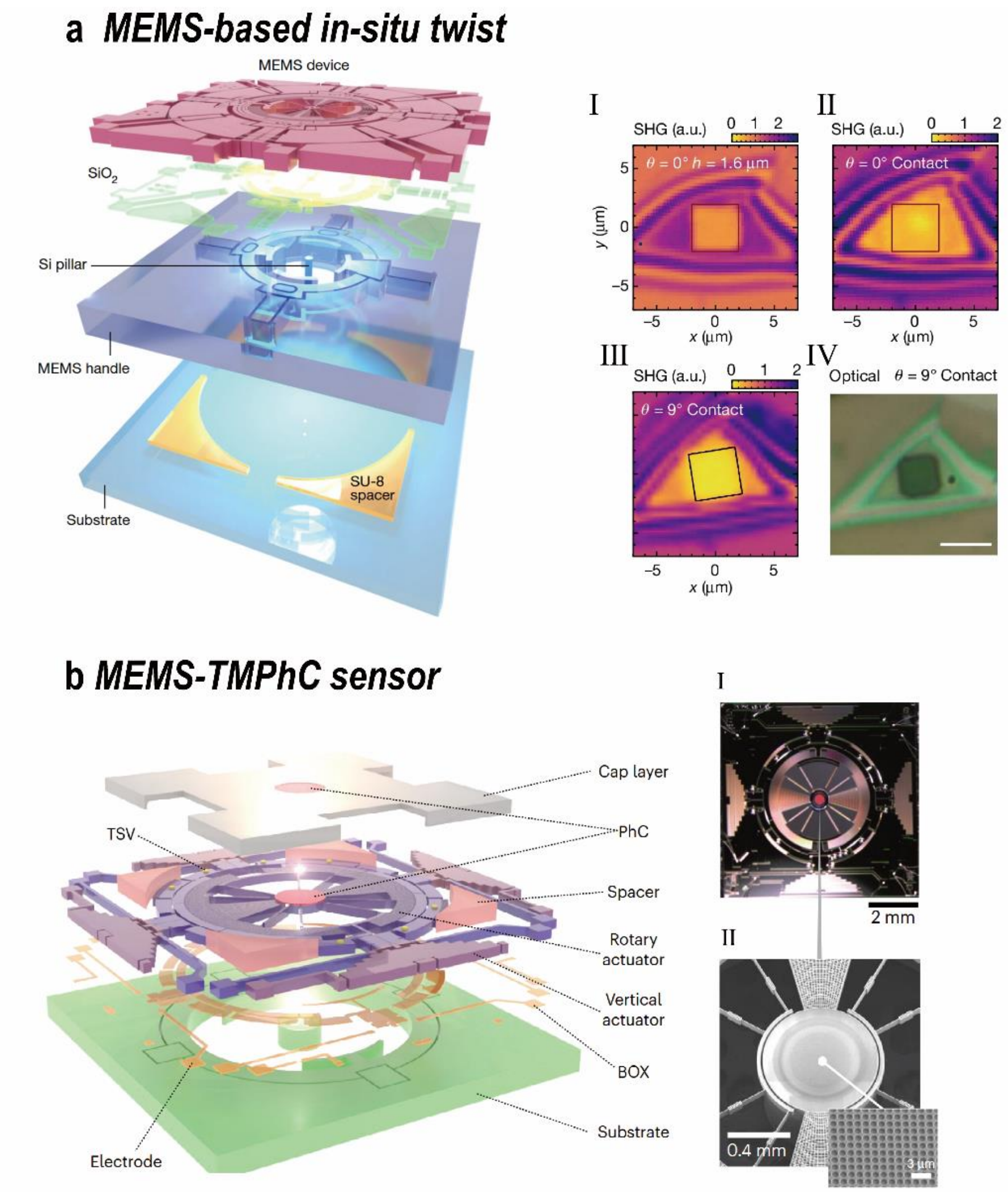

**Fig. 6.** MEMS twist system. **a** The MEMS-based twist protocol[81]. I-V. SHG power map of an as-fabricated MEGA2D device with different contact status. **b** Microelectromechanically tunable twisted moiré photonic crystal sensor[74]. I. Microscope image of the MEMS-TMPhC sensor. II. Scanning electron microscope image of the centre of the MEMS-TMPhC sensor, showing one of the integrated PhCs (inset).

Ultimately, the transformative role of MEMS will lie not only in enabling dynamic control, but also in establishing a standardized technological foundation for the broader moiré photonics community. As fabrication protocols become increasingly compatible with CMOS processes and wafer-scale manufacturing, MEMS platforms are expected to transition from bespoke research instruments to scalable enabling technologies for integrated photonic circuits. In this vision, MEMS will no longer serve merely as a high-precision actuator, but as the core infrastructure that bridges fundamental investigations of twist-dependent photonic phenomena with programmable, manufacturable, and application-ready moiré photonic systems.

### *4.3 Toward Closed-Loop Twist Engineering*

While current QTM implementations have already demonstrated unprecedented capability for momentum-resolved probing with continuously tunable twist angles ((Fig. 7a)[79] (Fig. 7b)[127]), their long-term significance extends well beyond improvements in angular precision or spectroscopic resolution. The true transformative potential of QTM lies in its ability to fundamentally merge measurement and twist control into a single experimental framework, thereby eliminating the conventional separation between device actuation and characterization. Rather than first fabricating a structure, subsequently measuring its properties, and then manually adjusting the twist angle, future QTM platforms are expected to enable simultaneous manipulation, characterization, and optimization of moiré systems in real time. Such a transition would establish QTM as a cornerstone technology for closed-loop twist engineering, where structural evolution and physical responses are continuously monitored and actively controlled within the same experimental cycle.

Looking forward, one important direction will be to reduce the technological barrier associated with QTM. At present, QTM remains a highly specialized technique that requires sophisticated probe fabrication, complex instrumentation, and expert operation, restricting its adoption to a limited number of research laboratories. Future developments should therefore prioritize standardized probe architectures, modular instrumentation, and user-friendly operating protocols, transforming QTM from a bespoke experimental setup into a broadly accessible research platform. Such advances would significantly accelerate the dissemination of momentum-resolved twist engineering across the wider moiré photonics and quantum materials communities.

Equally important, future QTM systems are expected to evolve from passive characterization tools into intelligent autonomous discovery platforms. By integrating ultrafast optical spectroscopy, Raman scattering, photoluminescence mapping, near-field optical microscopy, and machine-learning-assisted data analysis, QTM could continuously identify emergent optical states while autonomously adjusting the twist angle to maximize or stabilize targeted functionalities. Instead of manually scanning the twist-angle phase space, the system would actively search for photonic magic angles, topological transitions, or optimal coupling conditions through real-time feedback. In this vision, measurement is no longer a process performed after structural manipulation, but becomes an integral component of the manipulation itself.

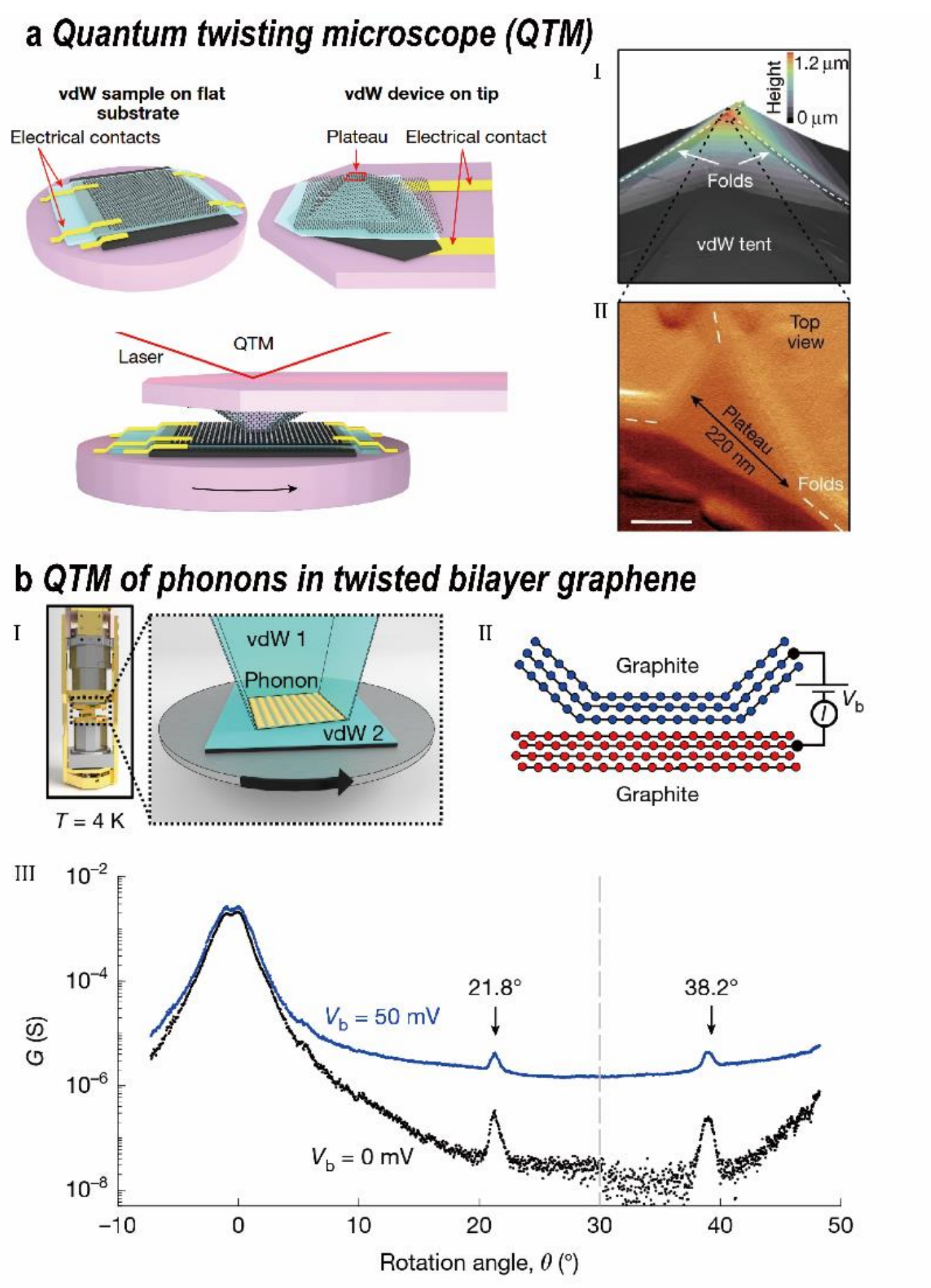

**Fig. 7.** QTM twist system. **a** The quantum twist microscope (QTM). The QTM can allow the in-situ twist and measurement of bi-layer graphene[79]. **b** Quantum twisting microscopy of phonons in twisted bilayer graphene[127]. I. Schematics of the cryogenic QTM (inset). II. Measure the phonon dispersion between the double-layer graphite sheets. III. Measured G versus twist angle, θ, for $V_b$ = 0 mV (black) and 50 mV (blue).

Ultimately, the future role of QTM will not be defined solely by its capability to observe moiré phenomena with exceptional precision, but by its evolution into a unified platform that seamlessly integrates measurement, actuation, feedback, and autonomous optimization. Such convergence represents a conceptual shift from conventional characterization toward self-driving experimentation, providing a powerful technological foundation for the discovery of previously inaccessible moiré photonic states and the realization of truly programmable quantum and photonic systems.

*4.4 Toward Wafer-Scale Programmable Moiré Photonics*

The ability to engineer twist angles has transformed van der Waals heterostructures into a versatile platform for exploring emergent electronic, optical, and quantum phenomena (Fig. 8)[133, 134]. Hoang et al. proposed a roll-to-roll nano-imprinting manufacturing scheme for visible light metasurfaces (Fig. 8a)[133]. By replacing the traditional expensive electroforming nickel molds with flexible polymers to replicate the molds and combining roll-to-roll ultraviolet nanoimprinting with atomic layer deposition of titanium dioxide coatings, they achieved large-scale, automated, and continuous fabrication of metasurfaces. Li et al. reported a scalable roll-to-roll printable meta-assembly approach for fabricating hierarchical photonic architectures over meter-scale areas (Fig.8b)[134]. The engineered coupling between guided and reflected optical modes enables programmable structural coloration with high efficiency and pixel-level control. This work demonstrates the potential of continuous manufacturing for large-area photonic metasurfaces and highlights a practical route toward scalable optical devices and multifunctional photonic systems. Yet, despite rapid progress in deterministic transfer and rotational alignment, twist-angle engineering remains predominantly a laboratory-scale endeavor. The fabrication of twisted bilayers still relies heavily on manual manipulation, limiting throughput, reproducibility, and scalability. Moving beyond proof-of-concept demonstrations will require a fundamental transition from handcrafted assembly to autonomous manufacturing. In our view, one of the most promising pathways toward this objective lies in the convergence of roll-to-roll (R2R) manufacturing, artificial intelligence, and real-time process control.

Future R2R platforms should evolve far beyond simply automating existing transfer procedures. Instead, they should function as intelligent manufacturing systems capable of continuously sensing, interpreting, and optimizing the assembly process. Machine vision combined with data-driven image analysis offers a promising route toward this goal by enabling rapid identification of crystal orientation, lattice symmetry, and interlayer rotational mismatch directly during transfer. When integrated with adaptive rotational stages and closed-loop feedback algorithms, such systems could autonomously compensate for alignment errors, structural relaxation, and process fluctuations in real time, transforming twist-angle alignment from a manually optimized operation into a self-correcting manufacturing process. More importantly, future manufacturing platforms should no longer treat twist angle as a predefined fabrication parameter, but as a dynamically optimized process variable that continuously evolves toward targeted device functionality.

Looking further ahead, the next generation of R2R manufacturing is expected to transition from geometry-driven fabrication to function-oriented manufacturing. Rather than producing structures with predetermined twist angles, intelligent production lines could directly optimize optical, electrical, or quantum functionalities through continuous feedback from in situ characterization. By integrating optical imaging, spectroscopic monitoring, electrical measurements, and machine-learning algorithms within a unified manufacturing framework, future systems may autonomously identify optimal moiré configurations, compensate for process variations, and adapt assembly strategies without human intervention. Such capability would establish a self-learning manufacturing paradigm in which device performance, rather than structural geometry alone, becomes the primary optimization target.

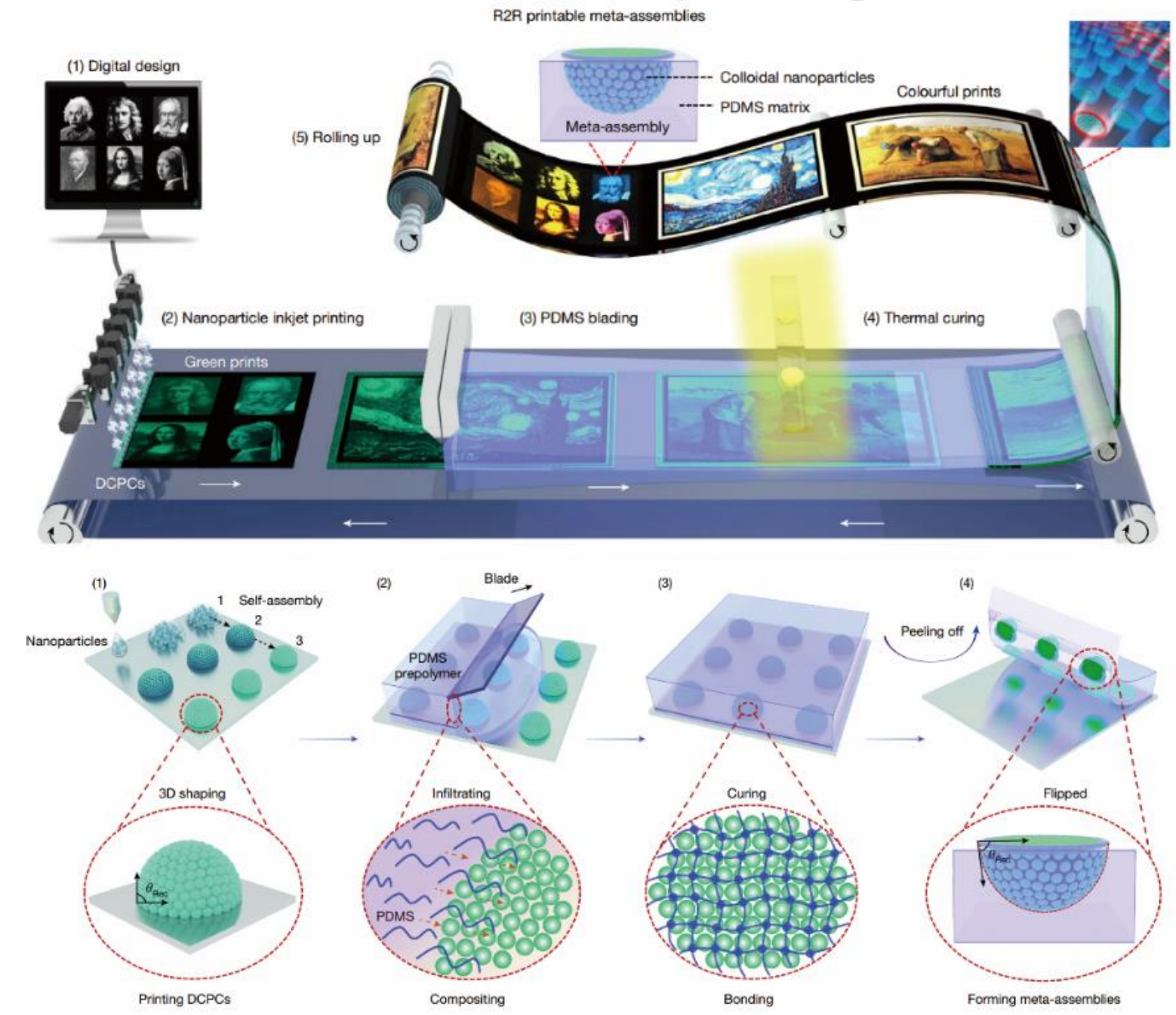

**Fig. 8** Roll-to-roll processing technology. **a** Mass manufacturing process of large-area metalens arrays[133]. **b** Overview of the printable meta-assembly strategy[134].

Equally important, future R2R technologies should emphasize standardization, interoperability, and compatibility with semiconductor manufacturing infrastructure. Developing modular process architectures, unified metrology protocols, and CMOS-compatible fabrication workflows will be essential for translating twisted van der Waals heterostructures from bespoke laboratory demonstrations to wafer-scale photonic manufacturing. Digital-twin technologies and predictive process models may further enable virtual optimization before physical assembly, substantially improving manufacturing yield and reducing process variability. In this vision, the production line itself becomes an intelligent experimental platform capable of continuously refining fabrication strategies through accumulated manufacturing data.

Ultimately, the significance of R2R manufacturing extends beyond increasing production throughput. Its transformative role lies in establishing an autonomous manufacturing ecosystem in which material synthesis, twist alignment, multimodal characterization, intelligent decision-making, and device optimization are seamlessly integrated within a closed-loop workflow. Such a paradigm would fundamentally redefine twist engineering from a manual fabrication strategy into a data-driven manufacturing science, paving the way for scalable, programmable, and industrially deployable moiré photonic technologies.

**5. Outlook and Challenges**

Twist engineering has rapidly evolved from a structural design strategy into a programmable platform for controlling electronic, optical, and quantum phenomena[38, 102, 135]. The emergence of moiré photonics signals a broader transformation in the way photonic systems are conceived, engineered, and ultimately manufactured. Rather than serving merely as a geometric descriptor of stacked structures, the twist angle is increasingly becoming a dynamically addressable degree of freedom that governs light–matter interactions across multiple spatial and temporal scales.

Looking forward, the future development of twisted bilayer photonic crystal membranes is likely to be driven by the coordinated evolution of several complementary technological directions. AFM-based platforms will remain indispensable for discovering previously inaccessible angle-dependent moiré phenomena and mapping the fundamental physical landscape[117, 132]. MEMS technologies will translate twist control into electrically programmable and highly integrated functionalities suitable for practical photonic devices[74, 81]. Quantum Twisting Microscopy (QTM) is expected to further bridge actuation and characterization through real-time, momentum-resolved feedback, enabling closed-loop exploration and optimization of complex moiré phase spaces[79, 127]. Ultimately, AI-enabled roll-to-roll manufacturing may provide the pathway toward wafer-scale, high-throughput production of programmable moiré photonic architectures[133, 134].

More broadly, the convergence of precision metrology, dynamic actuation, intelligent feedback, haptics, human machine interaction and scalable manufacturing points toward a conceptual shift in materials and device engineering[136-141]. Future twist engineering will no longer focus solely on constructing structures with predefined rotational configurations, but on continuously sensing, adapting, and optimizing interlayer configurations according to targeted functionalities. In such systems, the twist angle will cease to be a static fabrication parameter and instead become a dynamically controlled state variable that can be programmed, monitored, and optimized throughout the entire device lifecycle.

Perhaps most importantly, the next generation of moiré photonics is likely to move beyond manually designed and experimentally optimized devices toward autonomous discovery and intelligent manufacturing. By tightly integrating multimodal characterization, programmable actuation, machine learning, and digital manufacturing into a unified closed-loop framework, future platforms may autonomously identify optimal moiré configurations, discover previously unknown physical phenomena, and fabricate functional photonic devices with minimal human intervention.

In this emerging paradigm, twist will no longer represent merely an angle between two layers; it will become a programmable language through which photonic functionality is discovered, engineered, and manufactured.

**Acknowledgments**

This work was supported by the Agency for Science, Technology and Research (A*STAR) under MTC YIRG Grant No. M23M7c0129, Career Development Fund-Seed Projects Grant No. 222D800038, A*STAR Quantum Innovation Centre (Q. InC) SRTT. This work is also supported by the National Research Foundation, Singapore through the National Quantum Office, hosted in A*STAR, under its Centre for Quantum Technologies Funding Initiative (S24Q2d0009). Wu Lin acknowledges support from the National Research Foundation, Singapore (Grant Nos. NRF-CRP26-2021-0004 and NRF-CRP31-0007), the Ministry of Education, Singapore (Grant Nos. MOE-T2EP50223-0001, MOE-MOET32024-0005, and MOE-T2EP50125-0018), the Agency for Science, Technology and Research (A*STAR) (Grant No. MTC IRG M24N7c0083), and the Singapore University of Technology and Design (SUTD) under the Kickstarter Initiative (Grant

No. SKI 2021-04-12), the CAS Pioneer Hundred Talents Program and the National Natural Science Foundation of China (Grant No. 25XR532001).

## Author contributions

All authors contributed to the preparation of the manuscript.

## Competing interests

Authors declare that they have no competing interests.